\documentclass{article}

\usepackage[authoryear,round]{natbib}
\usepackage[hidelinks]{hyperref}

\usepackage{PRIMEarxiv}
\usepackage{amsmath} 
\usepackage{tabularx}
\usepackage{ragged2e}
\usepackage{array}
\newcolumntype{Y}{>{\RaggedRight\arraybackslash}X}
\usepackage[utf8]{inputenc} 
\usepackage[T1]{fontenc}    
\usepackage{hyperref}       
\usepackage{url}            
\usepackage{booktabs}       
\usepackage{amsfonts}       
\usepackage{nicefrac}       
\usepackage{microtype}      
\usepackage{lipsum}
\usepackage{fancyhdr}       
\usepackage{graphicx}       
\graphicspath{{media/}}     

\title{
From self-organizing systems to
subjective temporal extension
}

\author{
Jan Erik Bellingrath \\
Université de Toulouse \\
CerCo, CNRS \\
Toulouse, France\\ 
bellingrathjanerik@gmail.com
   \\
}

\usepackage{setspace}
\makeatletter
\AtBeginDocument{%
  \setstretch{1.9}%
  \renewcommand\normalsize{%
    \@setfontsize\normalsize{12pt}{10pt}%
    \abovedisplayskip 12pt plus 2pt minus 5pt%
    \abovedisplayshortskip \z@ plus 3pt%
    \belowdisplayshortskip 7pt plus 3pt minus 4pt%
    \belowdisplayskip \abovedisplayskip%
  }%
  \normalsize
}
\makeatother

\begin{document}
\maketitle

\begin{abstract}
The self-simulational theory of temporal extension (SST) describes an information-theoretically formalized 
mechanism by which the ‘width’ of subjective temporality emerges from the architecture of self-modelling. In this paper, the perspective of the free energy principle will be assumed to cast the 
emergence of subjective temporal extension from first principles of the physics of self-organization and to formalize subjective temporal extension using information geometry. Using active inference, a deep
parametric generative model of temporal inference is simulated, which realizes the described 
dynamics on a computational level. Two ‘biases’ (i.e., variations) of time-perception naturally emerge 
from the simulated computational model. This concerns the intentional binding effect (i.e., the 
compression of the temporal interval between voluntarily initiated actions and subsequent sensory 
consequences) and empirically documented alterations of subjective time experience in deep states 
of meditative absorption (i.e., in minimal phenomenal experience). Generally, numerous systematic
and domain-specific alterations of subjective temporal experience are computationally explained in a 
unified manner, as enabled by integration with current active inference accounts mapping onto the 
respective domains. This concerns, next to more general attentional and 
central tendency effects, the temporality-modulating role of valence, impulsivity, boredom, flow-states, near death-experiences, and various psychopathologies, amongst others. The self-simulational 
theory of temporal extension, from the perspective of the free energy principle, explains how the 'width' of the
subjective temporal moment emerges and varies from first principles, accounting for why sometimes, 
subjective time seems to fly, and sometimes, moments feel like eternities; with the computational 
mechanism being readily deployable synthetically.
\end{abstract}

\section{Introduction}

\begin{quote}
\textit{“Time is the substance I am made of. Time is a river which sweeps me along, but I am
the river.”} (\citeauthor{borges1964labyrinths}, 1962)
\end{quote}

Has time ever seemed to pass by too quickly? Have moments ever seemed like an
eternity? While the subjective experience of time seems as prevalent as the experience
of change, the experience of the subjective present is not simply the experience of
change, or change of change, etc. (\citealt{kent2021missinglink}). Our experience of subjective temporality is the experience of an extended moment, a specious present
(\citealt{james1890principlespsychology}), extending retrospectively to the moments that just happened, and
prospectively to the moments that are predicted to subsequently arise. Historically,
estimates for the ‘width’ of a subjective temporal moment range back over the
centuries, with the Buddhist Treasury of Abhidharma, for example, estimating there to be 65 temporal moments in the time it takes to snap one's fingers; with the ‘width’ of
the subjective temporal 'Now' thereby becoming about 1/65 of a second (i.e., about 15
milliseconds) (\citealt{thompson2014wakingdreamingbeing}). Essentially, physical time just \textit{is}, and does not extend in
the subjective sense. Accordingly, the ‘width’ of subjective temporality cannot be
perceptually inferred from external surroundings but must be the consequence of a \textit{counterfactual} inference
process (i.e., prospective or retrodictive inferences), entertained at a given moment of
physical time. 

Counterfactual self-simulations constitute one possibility for this
inferential process: Indeed, at least in the normal waking state, it looks as if there is always "someone"
flowing in the temporal river. Mirroring this phenomenology, numerous empirical
connections between temporal inference and self-modelling have been discovered. For
instance, signals from the heart cycle and timing responses in a duration reproduction
task are characterized by a certain synchrony (\citealt{pollatos2014askyourheart}).
Additionally, the amplitude of the heartbeat-evoked potential is associated with the
length of the estimated durations (\citealt{khoshnoud2024brainheart}; \citealt{richter2021timeisbody}); and
individuals who demonstrate a higher interoceptive accuracy (i.e., individuals that are
more accurate at counting the number of heartbeats during a given interval) show a
better performance in duration reproduction tasks (\citealt{meissner2011bodysignals}).

Consistent with these studies suggesting the existence of a system-relative (i.e.,
\textit{embodied}) temporal reference frame, neuroimaging meta-analyses reveal both self-modelling and temporal inference as centrally implicated in insula functioning (for
a meta-analysis on self-modelling, see e.g., \citealt*{qin2020threelevelself}; for meta-analyses on temporal inference, see \citealt{mondok2023selectivitytiming}; \citealt{naghibi2023embodyingtime}). This
is consistent with earlier proposals of the insula cortex as, on the one hand, the neuronal
substrate of subjective time perception (\citealt{craig2009emotionalmoments}) and the conscious feeling of
presence (\citealt{seth2012consciouspresence}); and, on the other hand, the neuronal substrate of
interoceptive inference and self-modelling (e.g., \citealt{seth2013interoceptiveinference}; \citealt{seth2016activeinteroceptive}).
Moreover, subjective time perception and the conscious feeling of presence have been
already, and by these very accounts, tied to the “material me” (\citealt{craig2009emotionalmoments}) and self-modelling and agency more generally (\citealt*{seth2012consciouspresence}). 

Psychologically,
furthermore, various normal (e.g., \citealt{rey2017paindilates}; \citealt*{witowska2020waiting}) and altered (for a review see \citealt{wittmann2015selftime}) states, traits, and
psychopathological conditions (e.g., \citealt*{kent2023transdiagnostic}; \citealt{martin2014temporalstructure})
are associated with covariant changes in self-modelling and temporal inference. These
states and traits include, as will be reviewed and explained below, amongst others,
varying levels of valence, impulsivity, boredom, the flow-state, near-death experiences,
the intentional-binding effect, and concentrated states of deep meditative absorption.
Consistently, many of these temporality-modulating constructs are related to functional
activation, and structural alterations, in the insula cortex (for a review in the context of
psychopathologies, for instance, see \citealt{vicario2020insula_minireview}). 

While this covariation, holding
across levels of description, methods, tasks, and theoretical perspectives, is too close
for a coincidence to be likely, a theory explicitly relating self-modelling and temporal
inference, in the sense of one being emergent from the other, has been lacking. The
self-simulational theory of temporal extension (SST) is a theory about the emergence of
subjective temporal extension (i.e., the 'width' of subjective temporality) from processes of self-modelling; formalized with
information-theory and formulated on a neuronally plausible level of description
(\citealt{bellingrath2023selfsimulational}). 

In this paper, the perspective of the
free energy principle (FEP) (\citealt{friston2010freeenergyprinciple}) will be assumed, to cast the emergence of
subjective temporal extension from first principles of the physics of self-organization.
The first part of this paper
will, after a section introducing the FEP and active inference, be concerned with an information-geometric and a computational description of the ‘width’
of subjective temporality. In the second part of this
paper, the described dynamics are simulated with a deep parametric generative model
of temporal inference. Two empirically documented ‘biases’ (i.e., variations) of time
experience will be shown to naturally emerge from the simulation of the computational
model. In the third part of this paper, empirical evidence on systematic variations of
temporal experience from diverse psychological states, traits, and psychopathological
conditions will be explained in a unified manner, based on active inference accounts
mapping onto the respective domains. This deployment of mutual constraints between
phenomenological, computational, neuronal and behavioural levels of description is
historically and methodologically in line with the neurophenomenological research
program outlined by Francisco Varela (\citealt{varela1996neurophenomenology}), which has recently, based upon
integration with the free energy principle and active inference, been revived in terms
of the nascent field of computational (neuro-) phenomenology (\citealt{sandvedsmith2024deepcomputational}; \citealt{ramstead2022generativepassages}). This improved version of the self-simulational theory of temporal extension (SST) (\citeauthor{bellingrath2023selfsimulational}, 2023) is mathematically more principled, empirically more meaningful and predictive, and theoretically more integrated with current models in the cognitive sciences; and can  be considered as SST 2.0.

\section{Introduction to the free energy principle and active inference}

If \textit{things} exist, what must they do? Specifically, how can self-organizing systems, or \textit{things}, “maintain
their states and form in the face of a constantly changing environment” (\citeauthor{friston2010freeenergyprinciple} \citeyear{friston2010freeenergyprinciple})? This question
is attempted to be answered by the free energy principle (FEP) (\citeauthor{friston2010freeenergyprinciple} \citeyear{friston2010freeenergyprinciple}; see also, \citeauthor{clark2013whatevernext} \citeyear{clark2013whatevernext}, \citeyear{clark2015surfinguncertainty}; \citeauthor{friston2023simpler} \citeyear{friston2023simpler}; \citeauthor{hohwy2013predictivemind} \citeyear{hohwy2013predictivemind}; \citeauthor{parr2022activeinferencebook} \citeyear{parr2022activeinferencebook}). Originally proposed within the context of theoretical neuroscience as a unified brain
theory (\citeauthor{friston2010freeenergyprinciple} \citeyear{friston2010freeenergyprinciple}), accounting for perception, attention, action, and numerous phenomena in
between, the explanatory scope of the FEP has been expanded to include every \textit{thing} (\citeauthor{friston2019particularphysics} \citeyear{friston2019particularphysics}).
But what does it mean to be a \textit{thing} in the first place?

A \textit{thing} can be characterized via the notion of a Markov blanket (\citeauthor{pearl1998graphicalmodels} \citeyear{pearl1998graphicalmodels}). A Markov blanket, $b$, is
the set of states that separates the internal parts of a system, $\mu$, from its surrounding environment, $x$.
Intuitively, for the brain, internal (neural) states are statistically insulated from external objects by
sensory receptors and muscles. This is a statement of conditional independence: Given knowledge of
the blanket, internal and external states are conditionally independent -- any influence that $\mu$ or $x$
have on one another is mediated by $b$ (\citeauthor{parr2019markovblankets} \citeyear{parr2019markovblankets}):
\begin{equation}
\mu \perp x \mid b 
\;\Longleftrightarrow\;
p(\mu, x \mid b) = p(\mu \mid b)\, p(x \mid b)
\end{equation}

This enables us to speak about both the system we seek to characterize, as well as its surrounding
environment, as entities, or \textit{things}, with independent existence. To remain in their characteristic low
entropy state, however, \textit{things} must, on average, minimize the model-conditional improbability of impinging
sensory data, known as surprise, $\Im(y)$ (\citeauthor{friston2010freeenergyprinciple} \citeyear{friston2010freeenergyprinciple}). This is because entropy, $H[P(y)]$, is the long-term average of surprise, $\mathbb{E}_{P(y)}[\Im(y)]$:
\begin{equation}
H[P(y)] 
= \mathbb{E}_{P(y)}\!\left[\Im(y)\right]
= -\,\mathbb{E}_{P(y)}\!\left[\ln P(y)\right]
\end{equation}

Mathematically, the minimization of surprise is equivalent to the maximization of model-evidence, 
$\ln P(y)$, enabling a formulation of systemic preservation in terms of Bayesian inference (i.e., self-
evidencing (\citeauthor{hohwy2016selfevidencing} \citeyear{hohwy2016selfevidencing})). However, as exact Bayesian inference is usually computationally
intractable, as there are many hidden states that all need marginalizing out, the intractable quantities
(the model-evidence, $\ln P(y)$, and the posterior probability, $P(x \mid y)$), are substituted with quantities
that approximate them (the variational free energy, $F[Q,y]$, and the approximate posterior, $Q(x)$
(i.e., the recognition distribution)) (\citeauthor{parr2022activeinferencebook} \citeyear{parr2022activeinferencebook}). Accordingly, the minimization of variational free
energy is equivalent to the maximization of model evidence, $\ln P(y)$, while simultaneously minimizing
the KL-divergence, $D_{\mathrm{KL}}[Q(x)\Vert P(x\mid y)]$, of the approximate posterior, $Q(x)$, from the posterior
probability, $P(x\mid y)$. Variational free energy, $F[Q,y]$, is equivalent to surprise, $- \ln P(y)$, if this
divergence is zero (i.e., free energy is an upper bound on surprise) (\citeauthor{parr2022activeinferencebook} \citeyear{parr2022activeinferencebook}):
\begin{equation}
F[Q,y]
= D_{\mathrm{KL}}\!\bigl(Q(x)\,\Vert\,P(x \mid y)\bigr)
- \ln P(y)
\end{equation}

This enables a perspective on the emergence of genuinely inferential architectures from nothing but
systemic preservation itself (e.g., \citeauthor{friston2017a_processtheory} \citeyear{friston2017a_processtheory}). For example, in active inference, the minimization
of $D_{\mathrm{KL}}[Q(x)\Vert P(x\mid y)]$ corresponds to perception and the maximization of $\ln P(y)$ corresponds to
action (i.e., selectively sampling and changing the world to conform to prior expectations about it)
(\citeauthor{parr2022activeinferencebook} \citeyear{parr2022activeinferencebook}). Considering a simple generative model of Bayesian perceptual inference (see Figure
1), the likelihood mapping from hidden causes to their outcomes, $P(o \mid s)$, is inverted using prior
beliefs, $D$, and sensory data, $o$, giving the perceptual inference of hidden causes given outcomes,
$P(s \mid o)$. The likelihood mapping, $A$ (i.e., $P(o \mid s)$), is equipped with a precision term, $\gamma_A$, by
exponentiating each element in the $i$th row and $j$th column of $A$ by $\gamma_A$ and normalizing (\citeauthor{parr2018precisionfalse} \citeyear{parr2018precisionfalse}). This precision-weighting is understood in terms of attentional processes (\citeauthor{feldman2010attentionuncertainty} \citeyear{feldman2010attentionuncertainty}). A higher precision-weighting maps onto an increased attentional resource-allocation, with
faster and more reliable inferences being resultant (e.g., \citeauthor{sandvedsmith2021metawareness} \citeyear{sandvedsmith2021metawareness}).

\begin{figure}
    \centering
    \includegraphics[width=0.5\linewidth]{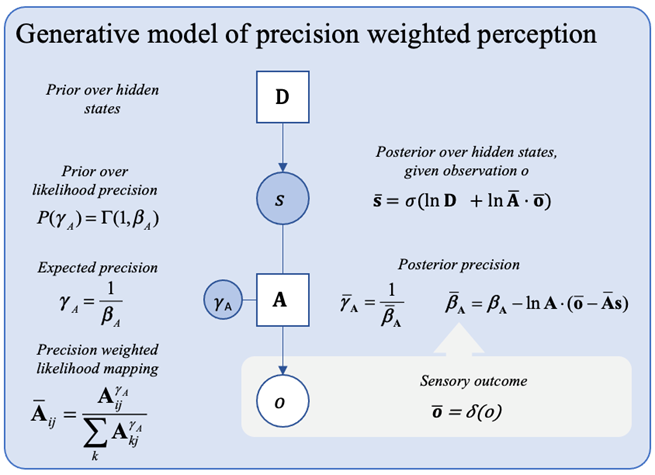}
    \caption{A probabilistic graphical model showing a basic generative model of precision-weighted perceptual inference. A
prior over hidden states is combined with sensory data to infer the posterior over hidden states given observations. Here,
Bayesian beliefs are noted in bold, bar notation represents posterior beliefs, $\sigma$ is the softmax function (returning a
normalized probability distribution), and $\delta$ is the Kronecker delta function (returning 1 for the observed outcome and zero
for all non-observed outcomes). The precision term, $\gamma_A$, over the likelihood mapping, $A$, is sampled from a gamma
distribution with inverse temperature parameter $\beta_A$. For the derivation of the precision belief update equation see (\citeauthor{parr2017uncertaintyepistemics} \citeyear{parr2017uncertaintyepistemics}) (supplementary appendix A.2); for the derivation of the latent state belief update equation, see (\citeauthor{friston2017b_processtheory} \citeyear{friston2017b_processtheory}) (supplementary appendix A). Figure and parts
of description reproduced with permission from (\citeauthor{sandvedsmith2021metawareness} \citeyear{sandvedsmith2021metawareness}); adapted from a template given in Figure 1a in
the study by (\citeauthor{hesp2021deeplyfeltaffect} \citeyear{hesp2021deeplyfeltaffect}).}
    \label{fig:placeholder}
\end{figure}

While variational free energy enables inferences about the future based on past data, it does not yet
account for prospective forms of inference (i.e., planning as inference (\citeauthor{botvinick2012planning} \citeyear{botvinick2012planning}))
over extended timescales based on anticipated future consequences (Figure 2). Counterfactual
simulations of sequences of actions (i.e., policies, $\pi$), associated with expected future observations
under those policies, are evaluated in terms of expected free energy, $G(\pi)$ (\citeauthor{friston2015epistemicvalue} \citeyear{friston2015epistemicvalue}).
Mathematically, the minimization of $G(\pi)$ is equivalent to the minimization of the divergence of the 
outcomes expected under a policy from the outcomes phenotypically expected (i.e., \textit{risk}), 
$D_{\mathrm{KL}}[Q(\tilde{y}\mid \pi)\Vert P(\tilde{y}\mid C)]$, while simultaneously minimizing \textit{ambiguity}, 
$\mathbb{E}_{Q(\tilde{x}\mid \pi)}[H[P(\tilde{y}\mid \tilde{x})]]$ (i.e., the 
expected inaccuracy due to an ambiguous mapping between states and outcomes) (\citeauthor{parr2022activeinferencebook} \citeyear{parr2022activeinferencebook}):
\begin{equation}
G(\pi)
= \mathbb{E}_{Q(\tilde{x}\mid \pi)}
\!\left[ H\!\left[P(\tilde{y}\mid \tilde{x})\right] \right]
+ D_{\mathrm{KL}}\!\bigl(
Q(\tilde{y}\mid \pi)\,\Vert\,P(\tilde{y}\mid C)
\bigr)
\end{equation}

$G(\pi)$ can also be decomposed in terms of \textit{epistemic} and \textit{pragmatic value,} with epistemic value, 
$\mathbb{E}_{Q(\tilde{y},\tilde{x}\mid \pi)}\big[D_{\mathrm{KL}}[Q(\tilde{x}\mid \tilde{y},\pi)\Vert Q(\tilde{x}\mid \pi)]\big]$, corresponding to the amount of information expected to be gained 
under the pursuit of a particular policy, and pragmatic value, $\mathbb{E}_{Q(\tilde{y}\mid \pi)}[\ln P(\tilde{y}\mid C)]$, being an expected 
utility:

\begin{equation}
G(\pi)
= -\,\mathbb{E}_{Q(\tilde{y},\tilde{x}\mid \pi)}
\!\left[
D_{\mathrm{KL}}\!\bigl(
Q(\tilde{x}\mid \tilde{y},\pi)
\,\Vert\,
Q(\tilde{x}\mid \pi)
\bigr)
\right]
- \mathbb{E}_{Q(\tilde{y}\mid \pi)}
\!\left[
\ln P(\tilde{y}\mid C)
\right]
\end{equation}

Policies evaluated in terms of low $G(\pi)$ are a priori more likely to 
be enacted. Equivalently, policies which have a high $G(\pi)$ are a priori surprising, which leads to their 
avoidance (\citeauthor{friston2015epistemicvalue} \citeyear{friston2015epistemicvalue}). Essentially, free energy minimizing systems come equipped with prior
preferences, $C$: These prior beliefs assign high probabilities to observations that are preferred to be
obtained. For example, human beings probabilistically expect their blood-glucose level to be in a
phenotypically beneficial range. By probabilistically presupposing prior preferences, $C$, and by then
acting to bring them about (as enabled by the selection of actions in terms of $G(\pi)$, influenced by $C$),
self-organizing systems self-evidence (\citeauthor{hohwy2016selfevidencing} \citeyear{hohwy2016selfevidencing}) as adaptive self-fulfilling prophecies (\citeauthor{friston2011embodiedinference} \citeyear{friston2011embodiedinference}). Selected policies are also influenced by a baseline prior over policies, $E$, which can be
understood in terms of the influence of habits (i.e., what the agent would do independent of $G(\pi)$).
For detailed introductions to active inference see (\citeauthor{parr2022activeinferencebook} \citeyear{parr2022activeinferencebook}; \citeauthor{smith2022activeinferencetutorial} \citeyear{smith2022activeinferencetutorial}; see also \citeauthor{friston2017a_processtheory} \citeyear{friston2017a_processtheory}).
\begin{figure}
    \centering
    \includegraphics[width=0.75\linewidth]{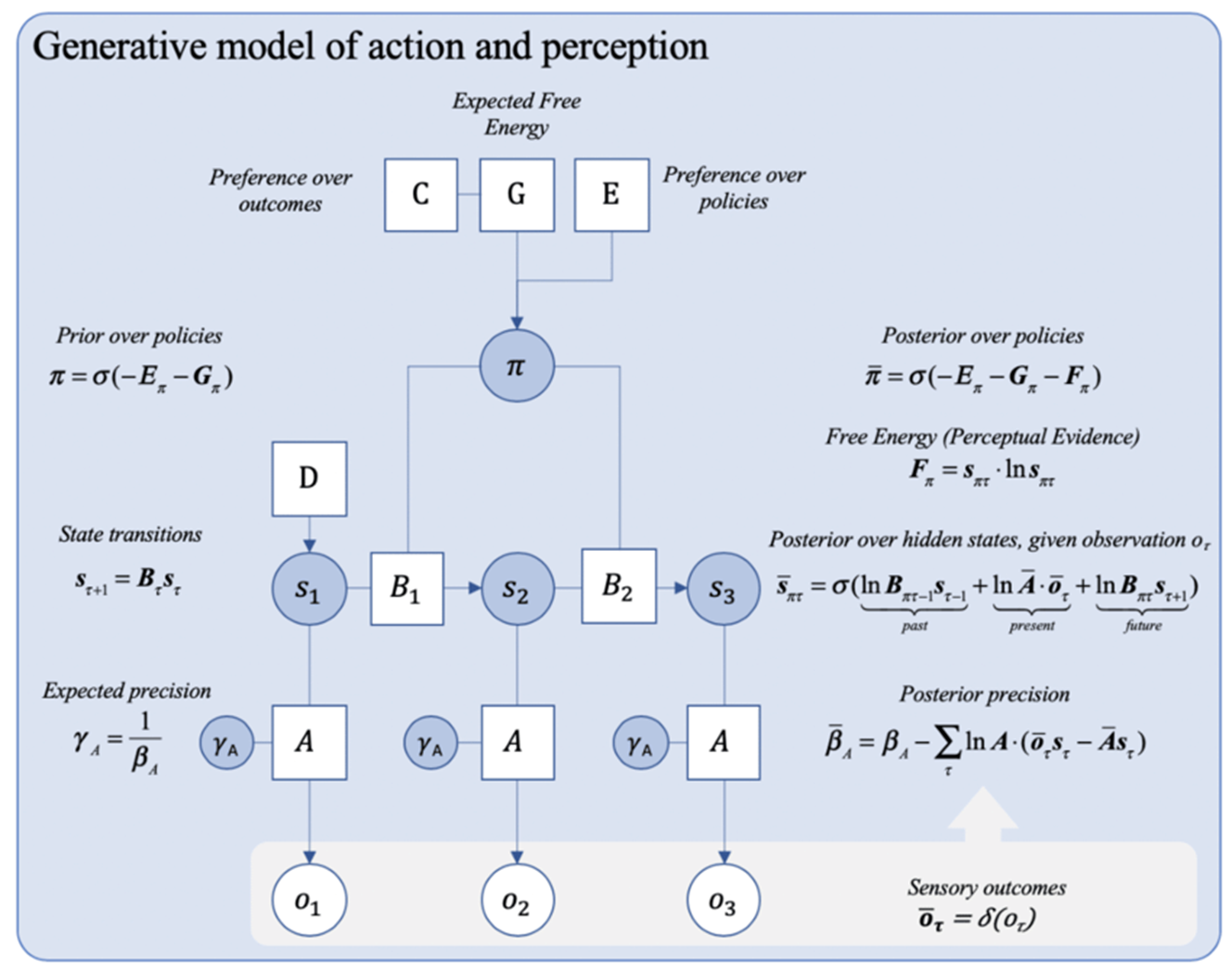}
    \caption{A probabilistic graphical model showing a deep generative model for policy selection. This model is equipped with beliefs about state transitions. Posterior state beliefs at each time step now depend on beliefs about the previous and subsequent states, mediated by the state transition matrix, B. (Figure and some parts of description reproduced with permission from (\citeauthor{sandvedsmith2021metawareness} \citeyear{sandvedsmith2021metawareness}); adapted from a template given in Figure 2 in the study by Hesp et al. (\citeauthor{hesp2021deeplyfeltaffect} \citeyear{hesp2021deeplyfeltaffect})).
}
    \label{fig:placeholder}
\end{figure}

The FEP and active inference present a parsimonious framework for computational phenomenology
(\citeauthor{sandvedsmith2024deepcomputational} \citeyear{sandvedsmith2024deepcomputational}; \citeauthor{ramstead2022generativepassages} \citeyear{ramstead2022generativepassages}). Indeed, many phenomenologically and
computationally central phenomena, such as self-modelling (\citeauthor{limanowski2013minimalselfmodels} \citeyear{limanowski2013minimalselfmodels}),
attention (\citeauthor{feldman2010attentionuncertainty} \citeyear{feldman2010attentionuncertainty}), valence (\citeauthor{hesp2021deeplyfeltaffect} \citeyear{hesp2021deeplyfeltaffect}), and meta-awareness (\citeauthor{sandvedsmith2021metawareness} \citeyear{sandvedsmith2021metawareness}), have already been understood in terms of the FEP and active inference. This
includes excellent work on aspects of subjective time experience (\citeauthor{albarracin2023mapping} \citeyear{albarracin2023mapping}; \citeauthor{bogota2023timeconsciousness} \citeyear{bogota2023timeconsciousness}; \citeauthor{hohwy2016distrustingpresent} \citeyear{hohwy2016distrustingpresent}; \citeauthor{wiese2017predictiveprocessingtime} \citeyear{wiese2017predictiveprocessingtime}).

\section{Subjective temporality as inference}

\begin{quote}
\textit{‘In this account one’s body is processed in a Bayesian manner as the most likely to be
“me”’} (\citeauthor{apps2014freeenergyself} \citeyear{apps2014freeenergyself})
\end{quote}

Self-modelling is a complex multidimensional process ranging across various
experiential domains, comprising interoceptive, emotional, and
bodily aspects, next to social, intentional, narrative and more broadly cognitive aspects
(e.g., \citeauthor{metzinger2003beingnoone} \citeyear{metzinger2003beingnoone}). The minimally sufficient form of phenomenal self-modelling is
known as minimal phenomenal selfhood (MPS) (\citeauthor{blanke2009fullbody} \citeyear{blanke2009fullbody}). MPS is the
simplest form of self-consciousness. In its original formulation, MPS comprised three
necessary and sufficient conditions: (1) global self-identification with the body as a
whole, (2) spatiotemporal self-location, and (3) a (weak) first-person perspective (\citeauthor{blanke2009fullbody} \citeyear{blanke2009fullbody}). In the following, and even though SST is also compatible with
MPS being a cluster-concept without a central individuating essence, current functional
level descriptions of self-modelling will be presented in the  context of MPS (for
a review on minimal self-modelling and the FEP, see (\citeauthor{limanowski2013minimalselfmodels} \citeyear{limanowski2013minimalselfmodels})). This will illustrate how subjective temporal extension emerges in the context
of the simplest forms of self-modelling.

Self-identification (with the body as a whole; the first condition of MPS) is the
consequence of a transparent inferential format (\citeauthor{blanke2009fullbody} \citeyear{blanke2009fullbody}; \citeauthor{metzinger2003beingnoone} \citeyear{metzinger2003beingnoone}). An inference is transparent if the system is unable to attend to the process of
construction of that inference. Precisely by not being able to introspectively notice that
a given inference is, in fact, constructed, the inference is imbued with a transparent (i.e.,
‘naïve realistic’) status. Conversely, if the system can direct its attention to the process
of construction of a given inference, that inference can be inspected opaquely as ‘just a
construct’ (i.e., ‘just a thought’) (\citeauthor{metzinger2003beingnoone} \citeyear{metzinger2003beingnoone}). The opacification of a transparent
inferential format is, in active inference, understood in terms of the deep parametric
architecture of hierarchical generative models (\citeauthor{limanowski2018seeingdark} \citeyear{limanowski2018seeingdark}; \citeauthor{sandvedsmith2021metawareness} \citeyear{sandvedsmith2021metawareness}), with ‘parametric depth’ referring to a nested architecture of ‘beliefs about beliefs’ (\citeauthor{hesp2021deeplyfeltaffect} \citeyear{hesp2021deeplyfeltaffect}). Here, the precision-weighting, $\gamma_A$, of the likelihood, $A$
(i.e., $P(o \mid s)$), corresponds to a central aspect of the process of construction of a given
probabilistic inference. Accordingly, the process of construction of a hierarchically lower
level probabilistic inference can be inspected by predicting its precision-weighting and
inferred state via an inference on a hierarchically higher level – with an opaque
inferential format of the hierarchically lower level inference being resultant (\citeauthor{limanowski2018seeingdark} \citeyear{limanowski2018seeingdark}; \citeauthor{sandvedsmith2021metawareness} \citeyear{sandvedsmith2021metawareness}). Transparency (i.e., ‘naïve realism’) is the
default inferential format (\citeauthor{metzinger2003beingnoone} \citeyear{metzinger2003beingnoone}), as evidenced both phenomenologically
and by the computational cost opacification (i.e., the additional presence of inferences
on a hierarchically higher level) engenders. Interestingly, complete intentional
opacification is logically impossible, because the highest-order inference is necessarily
transparent: its precision-weighting and inferred state would have to be predicted by
yet another higher-order prediction, which leads to an infinite regress of hierarchical
height (\citeauthor{limanowski2013minimalselfmodels} \citeyear{limanowski2013minimalselfmodels}). Expressed in the words of (\citeauthor{friston2012darkroom} \citeyear{friston2012darkroom}): “I can never conceive of what it is like to be me, because that
would require the number of recursions I can physically entertain, plus one”. While the
global self-identification with the body as a whole (the first condition of MPS) can be
understood in terms of a transparent inferential format of the body-representation
(\citeauthor{blanke2009fullbody} \citeyear{blanke2009fullbody}), empirical evidence from asomatic out-of-body
experiences and bodiless dreams suggests that a global self-identification with a body is
not a necessary feature of MPS, as in these states, self-identification occurs without a
global body-representation (e.g., self-identification as an extensionless point in space)
(\citeauthor{alcarazsanchez2022nothingness} \citeyear{alcarazsanchez2022nothingness}; \citeauthor{alvarado2000outofbody} \citeyear{alvarado2000outofbody}; \citeauthor{metzinger2013b_whydreams} \citeyear{metzinger2013b_whydreams}). 

The second condition in
the original formulation of MPS, spatiotemporal self-location (\citeauthor{blanke2009fullbody} \citeyear{blanke2009fullbody}), refers to a representation of a determinate volume in a spatial frame of
reference, normally localized within the bodily boundaries as represented, alongside the
represented ‘Now’ in a temporal frame of reference (from this conceptual perspective,
MPS presupposes (a minimal degree of) temporal ‘width’; with the latter, as will be
detailed below, emerging from \textit{subpersonal} (i.e., sub-MPS) dynamics of aspects of self-modelling, accordingly).

A weak first-person perspective, the third and last necessary
condition of MPS, is a purely geometric feature of self-modelling, which functions as the
geometric origin, or centre of projection, of the embodied system (\citeauthor{blanke2009fullbody} \citeyear{blanke2009fullbody}). While the weak first-person perspective was originally tied to perceptual
processing (e.g., visuospatial- or auditory-processing) (\citeauthor{blanke2009fullbody} \citeyear{blanke2009fullbody}), the
weak first-person perspective has also been understood completely independently of
perceptual processing (\citeauthor{windt2010immersive} \citeyear{windt2010immersive}).
Functionally, in terms of Bayesian inference, (\citeauthor{metzinger2013b_whydreams} \citeyear{metzinger2013b_whydreams}) identifies the origin of
the first-person perspective not with a specific inferential content, but rather with the
systemic region of maximal invariance. Indeed, the very existence of the agentic model
mandates the inclusion of the prior expectation that its form and internal states are
contained within some invariant set (\citeauthor{friston2011embodiedinference} \citeyear{friston2011embodiedinference}). As (\citeauthor{friston2011embodiedinference} \citeyear{friston2011embodiedinference}) writes: “This is
easy to see by considering the alternative: If the agent (model) entailed prior
expectations that it will change irreversibly, then (as an optimal model of itself), it will
cease to exist in its present state. Therefore, if the agent (model) exists, it must a priori
expect to occupy an invariant set of bounded states (cf., homeostasis)”. The
hierarchically highest (i.e., maximally invariant) inference, the belief that “I am a cause
in the world, or an agent” (\citeauthor{hohwy2017anybodyhaveself} \citeyear{hohwy2017anybodyhaveself}), is thereby equivalent to the
belief that “I act to maintain myself in certain states” (\citeauthor{hohwy2017anybodyhaveself} \citeyear{hohwy2017anybodyhaveself}), that is,
the prior preferences. As elaborated above, prior preferences, denoted by $C$, specify
phenotypically expected outcomes, $P(\tilde{y} \mid C)$, such as the expected blood-sugar level, or
the expected socio-emotional situatedness, and it is precisely the systemic
presupposition of the prior preferences that engenders their adaptive actualization akin
to a self-fulfilling prophecy (\citeauthor{friston2011embodiedinference} \citeyear{friston2011embodiedinference}). “Given the important links between the
notion of priors and the conditions that undergird an organism’s existence, we can also
say that in Active Inference, the identity of an agent is isomorphic with its priors” (\citeauthor{parr2022activeinferencebook} \citeyear{parr2022activeinferencebook}).

Self-modelling, however, is not unidimensional. In counterfactual inferences (i.e., in
prospectively predicting or retrodictively recalling), and for the transient time the simulated
episode persists, an aspect of our self-model is being simulated (\citeauthor{friston2018amiselfconscious} \citeyear{friston2018amiselfconscious};
\citeauthor{limanowski2020attenuating} \citeyear{limanowski2020attenuating}). This aspect of our self-model ‘lives through’ the
simulated action sequence, being confronted with the ‘outcomes that would result if I
were to act this way’ (i.e., $Q(\tilde{y} \mid \pi)$). This is the domain of mind-wandering (\citeauthor{smallwood2015mindwandering} \citeyear{smallwood2015mindwandering}), episodic memory (\citeauthor{tulving2002episodicmemory} \citeyear{tulving2002episodicmemory}), planning as inference (\citeauthor{botvinick2012planning} \citeyear{botvinick2012planning}), and related notions that centrally comprise self-relational
inferences in counterfactuality. Indeed, it has even been suggested that it is precisely
the change in the unit of identification that could serve to individuate individual episodes
of mind-wandering (\citeauthor{metzinger2013a_mythagency} \citeyear{metzinger2013a_mythagency}). Owing to the generality of the computational
problem that these counterfactual inferences (i.e., policies) are engendered to solve (i.e.,
minimization of expected free energy), policy-simulations are a virtually all-prevalent
aspect of our inferential architecture, being present over nested timescales, different
levels of hierarchical height, and in diverse inferential domains (e.g., \citeauthor{badcock2019hierarchically} \citeyear{badcock2019hierarchically}; \citeauthor{friston2017deeptemporal} \citeyear{friston2017deeptemporal}; \citeauthor{pezzulo2018motivatedcontrol} \citeyear{pezzulo2018motivatedcontrol}).

Motivationally, the outcomes expected under a counterfactual self-simulation, $Q(\tilde{y} \mid \pi)$,
are always evaluated in terms of a comparison (i.e., dissimilarity-relation) with
phenotypically preferred outcomes, $P(\tilde{y} \mid C)$. Intuitively, ‘what could’ is always
motivationally evaluated in terms of a comparison with ‘what ought’. In active inference,
this dissimilarity-relation between $Q(\tilde{y} \mid \pi)$ and $P(\tilde{y} \mid C)$ is expressed with the KL-divergence (Kullback and Leibler 1951), a central concept in information-theory. The
divergence of $Q(\tilde{y} \mid \pi)$ from $P(\tilde{y} \mid C)$, $D_{KL}[Q(\tilde{y} \mid \pi)\Vert P(\tilde{y} \mid C)]$, constitutes part of the
expected free energy, $G(\pi)$, decomposed in terms of risk (i.e., $D_{KL}[Q(\tilde{y} \mid \pi)\Vert P(\tilde{y} \mid C)]$)
and ambiguity (i.e., $\mathbb{E}_{Q(\tilde{x} \mid \pi)}[H[P(\tilde{y} \mid \tilde{x})]]$), in terms of which policies are evaluated (\citeauthor{parr2022activeinferencebook} \citeyear{parr2022activeinferencebook}):
\begin{equation}
G(\pi) = \mathbb{E}_{Q(\tilde{x} \mid \pi)}[H[P(\tilde{y} \mid \tilde{x})]] + D_{KL}[Q(\tilde{y} \mid \pi)\Vert P(\tilde{y} \mid C)]
\end{equation}
As elaborated above, $Q(\tilde{y} \mid \pi)$ and $P(\tilde{y} \mid C)$ are computational level descriptions of
aspects of self-modelling. Expressed from the transparent perspective of self-modelling
itself, $D_{KL}[Q(\tilde{y} \mid \pi)\Vert P(\tilde{y} \mid C)]$ corresponds to the ‘change that would have to happen for
(a part of) me to become that other (part of) me’. The central claim of the self-simulational theory of temporal extension is that this self-simulational dissimilarity relation is identical to the ‘width’ of subjective temporality: The subjective extension of
time is equivalent to the lived experience of the ‘length of the way towards another
version of yourself’.

\begin{equation}
\mathbb{TE}
\equiv
D_{\mathrm{KL}}\!\left[
Q(\tilde{y}\mid \pi)
\;\middle\Vert\;
P(\tilde{y}\mid C)
\right]
\qquad\text{(SST identity postulate)}
\end{equation}

Mathematically, by virtue of the properties of the KL-divergence, $\mathbb{TE}$ cannot be
negative (i.e., $D_{KL}[Q(\tilde{y} \mid \pi)\Vert P(\tilde{y} \mid C)] \ge 0$). Indeed, it is not even clear what a ‘negative
width’ of the subjective temporal Now could mean in principle. Crucially, physical time
just \textit{is}, and does not \textit{extend}, as subjective time does. Accordingly, because the ‘width’
of subjective temporality cannot be perceptually inferred, $\mathbb{TE}$ must be the
consequence of a counterfactual inference, entertained at a given moment of physical
time. Because $Q(\tilde{y} \mid \pi)$ (i.e., ‘what \textit{would} result, if I \textit{were} to act this way’) is a
counterfactual inference, it neither unfolds in the three-dimensional space available for
external action, nor in the \textit{n}-dimensional space available for internal (attentional) action.
However, a given counterfactual self-simulation could, if selected (in terms of $G(\pi)$) be
enacted in those spaces. Accordingly, because a counterfactual self-simulation could be
instantiated in the three- (and \textit{n}-) dimensional space of bodily and mental actions, and
because it is evaluated in terms of a dissimilarity-relation, $\mathbb{TE}$ (as part of $G(\pi)$), of how far it diverges
from this instantiation, it extends in a distinct dimension of self-simulational
counterfactuality, anchored on these spaces. Extension in a counterfactual dimension is
however simply what an inference of the ‘width’ of subjective temporality conceptually
presupposes – hence the proposed mechanism. $\mathbb{TE}$ varies: for some policies, $\mathbb{TE}$
might be very small, with subjective time being barely extended at all (i.e., subjective temporal
contraction). For other policies, $\mathbb{TE}$ might be larger, corresponding to a ‘wider’
temporal Now (i.e., subjective temporal dilation). 

Furthermore, subjective temporality, as
opposed to other inferences, such as the mental number line, is extended not only in
a cognitive, or opaque, sense, but in an immediate, \textit{transparent}, felt sense (\citeauthor{metzinger2003beingnoone} \citeyear{metzinger2003beingnoone};
\citeauthor{seth2012consciouspresence} \citeyear{seth2012consciouspresence}). Essentially, transparency is the default inferential format, being
both phenomenologically evident and entailed by the computational cost opacification
(i.e., the additional presence of inferences on a hierarchically higher level) engenders
(\citeauthor{limanowski2018seeingdark} \citeyear{limanowski2018seeingdark}; \citeauthor{metzinger2003beingnoone} \citeyear{metzinger2003beingnoone}). Accordingly, it is precisely because
introspective attention is not turned towards the process of construction of the
subjective temporal Now, that we simply \textit{live within it} without realizing its inferential
nature – with its extension varying systematically across our experience. 

Using the tools of information-geometry (e.g., \citeauthor{amari2016information} \citeyear{amari2016information}), the asymmetric KL divergence can, assuming infinitesimally small changes of the parameters, $\theta$, be
interpreted in terms of the symmetric Fisher information, $\mathcal{I}(\theta)$, a Riemannian metric
tensor of a statistical manifold, $\mathcal{M}$ (i.e., a space of probability-distributions). This is possible because by measuring the divergence over an
infinitesimally small change, $d\theta$, a second-order Taylor series expansion becomes
sufficient for its characterization; and $\mathcal{I}(\theta)$ is the local Hessian matrix of the KL-divergence
(Da Costa and Friston 2019; \citeauthor{friston2019particularphysics} \citeyear{friston2019particularphysics}). After integrating over a path in $\mathcal{M}$
$\Big(\text{with } p(\tilde y\mid\theta(t_0))=P(\tilde y\mid C),\;p(\tilde y\mid\theta(t_1))=Q(\tilde y\mid\pi)\Big)$, we obtain
the path length, traversed by neuronal dynamics, in the statistical manifold, thus yielding a distinct information geometric formulation of subjective temporal extension (locally proportional to the divergence and globally distinct). We can take a derivative with respect to
(external) time, thus deriving an instantaneous
relative measure of subjective and objective time, $\mathcal{R}(t)$:
\begin{equation}
D_{KL}\!\left[p(\tilde y \mid \theta)\,\Vert\,p(\tilde y \mid \theta + d\theta)\right]
\;\approx\;
\frac{1}{2}\, d\theta^{T}\mathcal{I}(\theta)\,d\theta
\end{equation}

\begin{equation}
\mathcal{I}(\theta)
=
-\,\mathbb{E}_{p(\tilde y \mid \theta)}\!\left[
\frac{\partial^{2}\log p(\tilde y;\theta)}{\partial \theta_i \partial \theta_j}
\right]
\end{equation}

\begin{equation}
l = \int_{t_0}^{t}
\sqrt{\left(\frac{d\theta}{dt'}\right)^{T}\mathcal{I}(\theta(t'))\left(\frac{d\theta}{dt'}\right)}
\, dt'
\end{equation}

\begin{equation}
\mathcal{R}(t) = \frac{dl}{dt}
\end{equation}

 Explicit Bayesian duration inference can be cast as a hierarchically higher level which generates top-down predictions about the expected magnitude of the self-simulational dissimilarity-relation. The mismatch is passed forward to update the duration inferences, thereby minimizing that error. Durations are overestimated following temporal moments that are very
extended, or ‘wide’, and underestimated for ‘shallower’, or fewer, temporal moments.

The Bayesian nature of this inference-scheme directly enables an account of Vierordt’s
law (the central-tendency effect): Participants in temporal reproduction tasks
overproduce shorter durations and underproduce longer durations (for reviews, see \citeauthor{gu2011vierordtslaw} \citeyear{gu2011vierordtslaw}; \citeauthor{lejeune2009vierordtlegacy} \citeyear{lejeune2009vierordtlegacy}). In line with Bayesian frameworks of timing
generally (e.g., \citeauthor{shi2013bayesianoptimizationtime} \citeyear{shi2013bayesianoptimizationtime}), this central-tendency effect can be
explained by the influence of the (acquired) hierarchically higher level duration
inference priors – with duration estimates being drawn towards previously inferred
interval-lengths. For an example of a deep
generative model equipped with duration inference states see Figure 3.

Generally, empirically, aspects of self-modelling and temporal inferences covary too systematically
for a coincidence to be likely (see sections 1, 4 \& 5). The self-simulational theory of
temporal extension explains this close covariation: subjective temporal extension emerges from
self-modelling. Put metaphorically, the ‘width’ of the subjective temporal moment is the
‘length of the way towards another version of yourself’. The observations that are
expected if a sequence of actions were enacted, $Q(\tilde{y}\mid \pi)$, are evaluated relative to the
observations that are phenotypically preferred, $P(\tilde{y}\mid C)$; with subjective time experience
emerging from their dissimilarity-relation implicated by the never-ending
phenomenological pursuit of ‘that other me, where I would have finally reached my
goal’.

\begin{figure}
    \centering
    \includegraphics[width=1\linewidth]{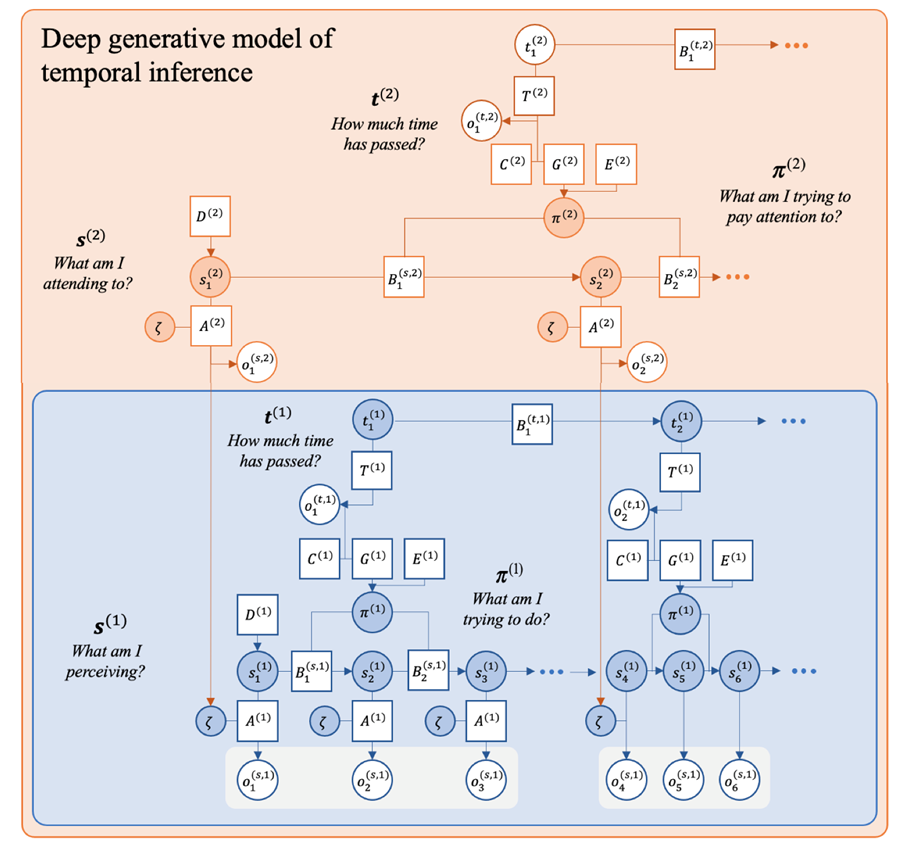}
    \caption{A probabilistic graphical model depicting an example of a deep parametric generative model with two hierarchical levels (perceptual and attentional). Each hierarchical level is equipped with duration inference states. A more extensive description of this kind of model-architecture can be found below. Adopted from a template given by (Sandved-Smith et al. 2021).}
    \label{fig:placeholder}
\end{figure}

\section{A deep parametric generative model of temporal inference}

The following simulation of a deep parametric generative model will serve to illustrate
the computational dynamics of the ‘width’ of subjective temporality, demonstrating
how two ‘biases’ (i.e., variations) of temporal inference naturally emerge from this
minimal model. The model-architecture features two modifications from the deep
generative model of mental action (\citeauthor{sandvedsmith2021metawareness}, \citeyear{sandvedsmith2021metawareness}), from which the use of
the oddball-perceptual task (consisting of the presentation of either a “standard” or a
“deviant” stimulus, with the former being presented 24/25 and the latter 1/25 times)
has also been adopted. The generative model is composed of different hierarchical
levels. The lowest layer of the model is concerned with perceptual inference,
specifically, with inferring whether the presented stimulus is a deviant (i.e., oddball)
stimulus or not, being equipped with a state-transition matrix $B^{(1)}$. As described above,
the precision-weighting, $\gamma_A$, of the likelihood, $A$, is understood in terms of attention
(\citeauthor{feldman2010attentionuncertainty}, \citeyear{feldman2010attentionuncertainty}). In this generative model, the precision-weighting of the
likelihood of the perceptual inference level is conditioned by hierarchically higher level
attentional states, which are in turn conditioned by attentional policies (\citeauthor{sandvedsmith2021metawareness}, \citeyear{sandvedsmith2021metawareness}). This is in accord with work on attentional control as precision deployment
(\citeauthor{brown2011activeinferenceattention}, \citeyear{brown2011activeinferenceattention}; \citeauthor{kanai2015pulvinarprecision}, \citeyear{kanai2015pulvinarprecision}). Faster and more reliable
perceptual inferences are induced by a higher precision-weighting. The hierarchically higher level attentional states are “focused” and “distracted”. These attentional states are, just as
the perceptual states, inferred via variational inference and equipped with a state transition matrix, $B^{(2)}$. The attentional actions on the second hierarchical level are either
“stay” or “switch”. “Stay” leads the agent to remain in its current attentional
state and “switch” flips the attentional state. The attentional actions are chosen from
attentional policies (also comprising “stay” or “switch”) based on their expected free
energy, $G(\pi)$. Actions are exclusively defined on this second hierarchical level,
conditioning the state-transition matrix $B^{(2)}$. The prior preferences over
attentional states, $C^{(2)}$, are set such that the agent prefers being in the state “focused”. This
generative model was originally deployed as a generative model of mental
action, modelling the dynamic attentional cycle in meditative practices (for a more
detailed description see \citeauthor{sandvedsmith2021metawareness}, \citeyear{sandvedsmith2021metawareness}). 

The original model architecture
(\citeauthor{sandvedsmith2021metawareness}, \citeyear{sandvedsmith2021metawareness}) has been modified in two distinct ways. These changes
concern (1) the sequentialization of policy-simulation, and (2) the incorporation of
resource-sensitive habitual control mechanisms. Specifically, while in the original model,
the two distinct policies (“stay” and “switch”) were both simulated at each time-step,
here, policy-simulation is sequentially unfolded in time, such that, at each time-step,
only one policy is (at most) simulated. Consistent with previous extensions of active
inference to resource-sensitive habitual control (\citeauthor{maisto2019cachinghabits}, \citeyear{maisto2019cachinghabits}),
policy-simulation is entirely absent at some time-steps. Specifically, first, if, at the end
of a simulation epoch (comprising the two policies “stay” and “switch” once, each), a
policy is evaluated in terms of a sufficiently low $G(\pi)$ (and thereby deemed sufficiently
‘good’), the next simulation epoch is skipped entirely, with the selected policy being
enacted for the next two time-steps. Intuitively, ‘once a sufficiently good plan has been
found, there is no need to reconsider for a while’. Second, if during each simulation
epoch (the order of the two policies is alternated), the $G(\pi)$ of the first simulated policy
is sufficiently low (and, more specifically, also below the $G(\pi)$ of both policies from the
previous simulation epoch), the rest of the simulation epoch (i.e., the other policy) is
skipped, with the first policy being immediately enacted. This, too, amounts to the
implicit assumption that contingencies do not change significantly in the absence of
policy-simulations (as deliberative control would be more advantageous otherwise
(\citeauthor{maisto2019cachinghabits}, \citeyear{maisto2019cachinghabits})), and can also be understood as the agent refraining from additional policy simulation, given that a sufficiently adaptive policy has already been
found, thereby saving significant resources. This is consistent with current perspectives
on the importance of resource-control for brain functioning (e.g., \citeauthor{barrett2016allostasis}, \citeyear{barrett2016allostasis}). From another perspective, this can be considered as simply another
consequence of free-energy minimization: Analogously to model-selection, non-necessary computations are avoided by the removal of excessive parameters
(\citeauthor{fitzgerald2014modelaveraging}, \citeyear{fitzgerald2014modelaveraging}; \citeauthor{friston2017a_processtheory}, \citeyear{friston2017a_processtheory}; \citeauthor{maisto2019cachinghabits}, \citeyear{maisto2019cachinghabits}; \citeauthor{pezzulo2015homeostatic}, \citeyear{pezzulo2015homeostatic}). Furthermore, the decreased frequency of (sequential) policy simulation on the hierarchically higher level, with selected policies thereby being (on
average) enacted for a longer time, mirrors empirical evidence on the increase of
intrinsic neuronal timescales with the hierarchical height of the cortex
(\citeauthor{friston2017deeptemporal}, \citeyear{friston2017deeptemporal};  \citeauthor{kiebel2008hierarchy}, \citeyear{kiebel2008hierarchy}; \citeauthor{murray2014intrinsictimescales}, \citeyear{murray2014intrinsictimescales}).

Strikingly, given the minimality of this generative model, two ‘biases’ (i.e., variations) of
temporal inference naturally emerge from the simulation. Well-known in cognitive psychology, the intentional binding effect (sometimes known as the temporal binding effect)
refers to the compression of the temporal interval between voluntarily initiated actions
and subsequent sensory consequences; often measured in terms of a direct interval
estimation procedure (for a review see \citeauthor{moore2012intentionalbinding}, \citeyear{moore2012intentionalbinding}). Indeed, this covariation
of temporal inference and self-modelling has been observed so reliably, that intentional
binding has been proposed as an implicit measure for the sense of agency
(\citeauthor{haggard2002actionbindingchapter}, \citeyear{haggard2002actionbindingchapter}; \citeauthor{haggard2002voluntaryaction}, \citeyear{haggard2002voluntaryaction}; though see \citeauthor{suzuki2019}, \citeyear{suzuki2019} for another perspective on intentional binding in terms of multisensory causal
binding). As elaborated above, and due to the free-energy minimizing resource-sensitive
habitual control mechanism, if a policy is evaluated in terms of a sufficiently low $G(\pi)$, the rest of
the simulation epoch (comprising the (alternating) second policy, “stay” or “switch”) is
skipped. $\mathbb{TE}$ mathematically constitutes, together with ambiguity, expected
free energy, $G(\pi)$, in terms of which the simulated policies are evaluated. Accordingly,
given the low $G(\pi)$ (implying a contracted $\mathbb{TE}$) of the selected policy, and the absence
of policy-simulation (and -evaluation) for the rest of the epoch, the temporal interval
between voluntarily initiated actions and resulting sensory consequences is
underestimated, explaining the intentional binding effect. The emergence of the intentional
binding effect is demonstrated in Figure 6 in terms of an increased number of
“congruent” episodes – episodes where the simulation and selection of a policy occur at
the same time-step. This congruence implies, on average, and because the simulated policy
is ‘already good enough’, a relatively decreased $\mathbb{TE}$ and relatively fewer subsequent
policy simulations. In contrast, “incongruent” episodes (here, a policy is not selected
immediately upon its simulation) imply, because of the relatively increased $G(\pi)$, an
increased $\mathbb{TE}$, and more subsequent policy simulations.

This generative model enables a computational account of alternating episodes of
concentration and distraction (\citeauthor{sandvedsmith2021metawareness}, \citeyear{sandvedsmith2021metawareness}), characteristic of
mindfulness meditation: Starting in a state in which the agent both is focused and infers itself to be focused, the agent, reflecting the intrusion of
all-too-frequent mind-wandering-episodes (\citeauthor{schooler2011metawareness}, \citeyear{schooler2011metawareness}; \citeauthor{smallwood2015mindwandering}, \citeyear{smallwood2015mindwandering}), becomes distracted without inferring this very fact. Then, after a period
of being distracted without having noticed, the agent infers that it is distracted, leading
to a reallocation of attentional focus (Figure 6). Mindfulness meditation has traditionally
(e.g., \citeauthor{analayo2003satipatthana}, \citeyear{analayo2003satipatthana}) and more recently been associated with attenuations of self-referential processing (\citeauthor{dahl2015reconstructing}, \citeyear{dahl2015reconstructing}; \citeauthor{milliere2020selflessness}, \citeyear{milliere2020selflessness}; \citeauthor{tang2015mindfulness}, \citeyear{tang2015mindfulness}), being mirrored on the neuronal level of description in terms of a reduced
activity of a network of regions implicated in mind-wandering (e.g., \citeauthor{brewer2011meditation}, \citeyear{brewer2011meditation}). From the perspective of active inference, the effects
of meditation, and ‘selfless’ experiences generally, have been understood in terms of a
reduced frequency and a reduced counterfactual depth of policy-simulations (i.e., self-simulations) (\citeauthor{laukkonen2021manytoone}, \citeyear{laukkonen2021manytoone}; \citeauthor{limanowski2020attenuating}, \citeyear{limanowski2020attenuating}; see also
Deane, Miller, and Wilkinson 2020).

In line with a rich tradition in eastern philosophy, alterations of temporal experience are
attested in deep states of meditative absorption (\citeauthor{metzinger2024elephantblind}, \citeyear{metzinger2024elephantblind}): A recent factor-analytic analysis of 1403 qualitative reports from meditators, directed at the experience
of ‘pure awareness’, an entirely non-conceptual and contentless form of awareness \textit{as
such} (i.e., awareness of nothing but awareness), converged on a 12-factor model to
describe the phenomenological character of these experiences in a fine-grained way
(\citeauthor{gamma2021mpe92m}, \citeyear{gamma2021mpe92m}). The factor that explained most of the variance (‘Time,
Effort and Desire’) integrated goal-directedness, effort and temporal experience.
Describing the phenomenological profile of this factor in meditative experience,
(\citeauthor{gamma2021mpe92m}, \citeyear{gamma2021mpe92m}) write: “Typically, there will be an attentional lapse,
followed by the phenomenology of noticing, remembering the goal state and refocusing. As a result, temporal experience is preserved: for example, we find the
phenomenology of duration and of time passing”. This factor is positively correlated
with mind-wandering, memories, the simulation of future events, and the arising of
thoughts and feelings generally (Factor 7: ‘Thoughts and Feelings’). This association of
temporal experience with goal-directed counterfactual inferences is also evident in the
generative model: The ‘width’ of subjective temporality (i.e., $\mathbb{TE} =
D_{KL}[Q(\tilde{y}\mid \pi)\Vert P(\tilde{y}\mid C)]$) varies characteristically, dependent on the counterfactual
observations expected under a given policy, $Q(\tilde{y}\mid \pi)$, evaluated in terms of systemic goal
states, $P(\tilde{y}\mid C)$. 

Strikingly, evidence from these qualitative reports attests to alterations
of temporal experience that culminate in phenomenal states that are
described as \textit{atemporal} (\citeauthor{metzinger2024elephantblind}, \citeyear{metzinger2024elephantblind}). However, because these states nevertheless
often feature experiences of change, it may be asked: “How can a physical system like
the human brain create conscious models of reality that, on a conceptual level, seem to
necessitate paradoxical descriptions like ‘timeless change’?” (\citeauthor{metzinger2024elephantblind}, \citeyear{metzinger2024elephantblind}). As can
be seen in Figure 6 of the generative model of temporal inference, perceptual states are
inferred at each time-step, with the perceptual state changing over time. However,
while the perceptual state may change, policies are not computed at each time-step.
Crucially, because the existence of $\mathbb{TE}$ depends on the simulation of a policy
(specifically, its evaluation in terms of $G(\pi)$), $\mathbb{TE}$ is entirely absent if no policy is
computed. In the generative model, furthermore, and consistent with the target
phenomenon, the probability for policies to not be computed increases in the focused
attentional state. As elaborated above, this reduced frequency of policy-simulation is
consistent with active inference accounts of the effects of meditation and ‘selfless’
experiences generally (\citeauthor{laukkonen2021manytoone}, \citeyear{laukkonen2021manytoone}; \citeauthor{limanowski2020attenuating}, \citeyear{limanowski2020attenuating}).
Furthermore, the absence of deliberative control (i.e., the absence of policy-simulation),
and the concomitant absence of a feeling of effort (Factor 1 integrated effort and time
(\citeauthor{gamma2021mpe92m}, \citeyear{gamma2021mpe92m})), is consistent with an active inference account of effort
– with effort being constitutively dependent on the presence of policy-simulations (i.e.,
effort being absent in the absence of policy-simulations) (\citeauthor{parr2023cognitiveeffort}, \citeyear{parr2023cognitiveeffort}). In
actual neuronal systems, a drastic reduction in the frequency of policy-simulation over
hierarchical levels might suffice for explaining the phenomenological reports (i.e., the
binary presence/absence distinction in the generative model is an idealization).
Accordingly, as the computation of counterfactual policies recedes more and more, the
subjective temporal Now ceases to be constructed. A continuous flow of perceptual
states unfolds in an \textit{effortless} and \textit{atemporal} phenomenal state: Dynamic succession
without temporal extension – the \textit{specious present }(\citeauthor{james1890principlespsychology}, \citeyear{james1890principlespsychology}) is not specious
anymore.

\begin{quote}
\textit{“The early Dzogchen scholar–practitioners in Tibet knew all of this very well, but
through their own meditation practice: “Nowness” is empty”} (\citeauthor{metzinger2024elephantblind}, \citeyear{metzinger2024elephantblind})
\end{quote}

\begin{figure}
    \centering
    \includegraphics[width=1\linewidth]{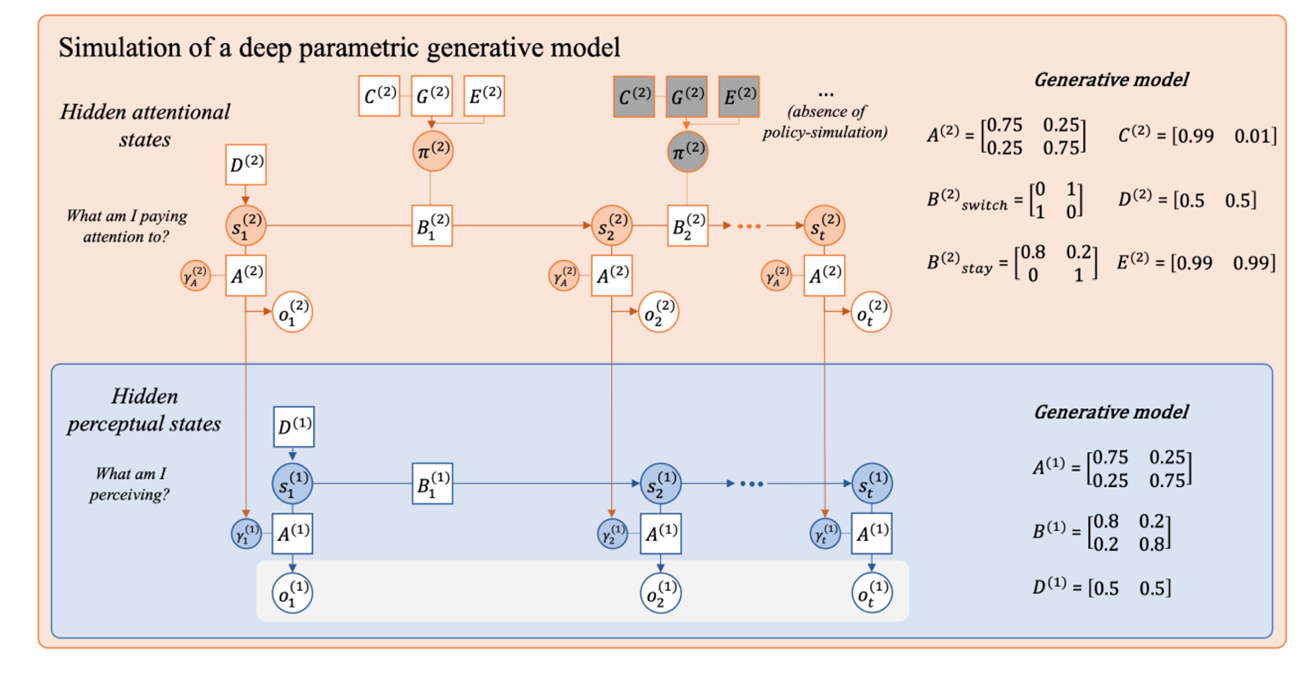}
    \caption{The simulated deep parametric generative model. Policies are only defined on the second hierarchical level. The
simulation of policies, and their evaluation in terms of expected free energy, is omitted at some time-steps (see the text).
Adopted from a template given by (\citeauthor{sandvedsmith2021metawareness}, \citeyear{sandvedsmith2021metawareness}).}
    \label{fig:placeholder}
\end{figure}

\begin{figure}
    \centering
    \includegraphics[width=1\linewidth]{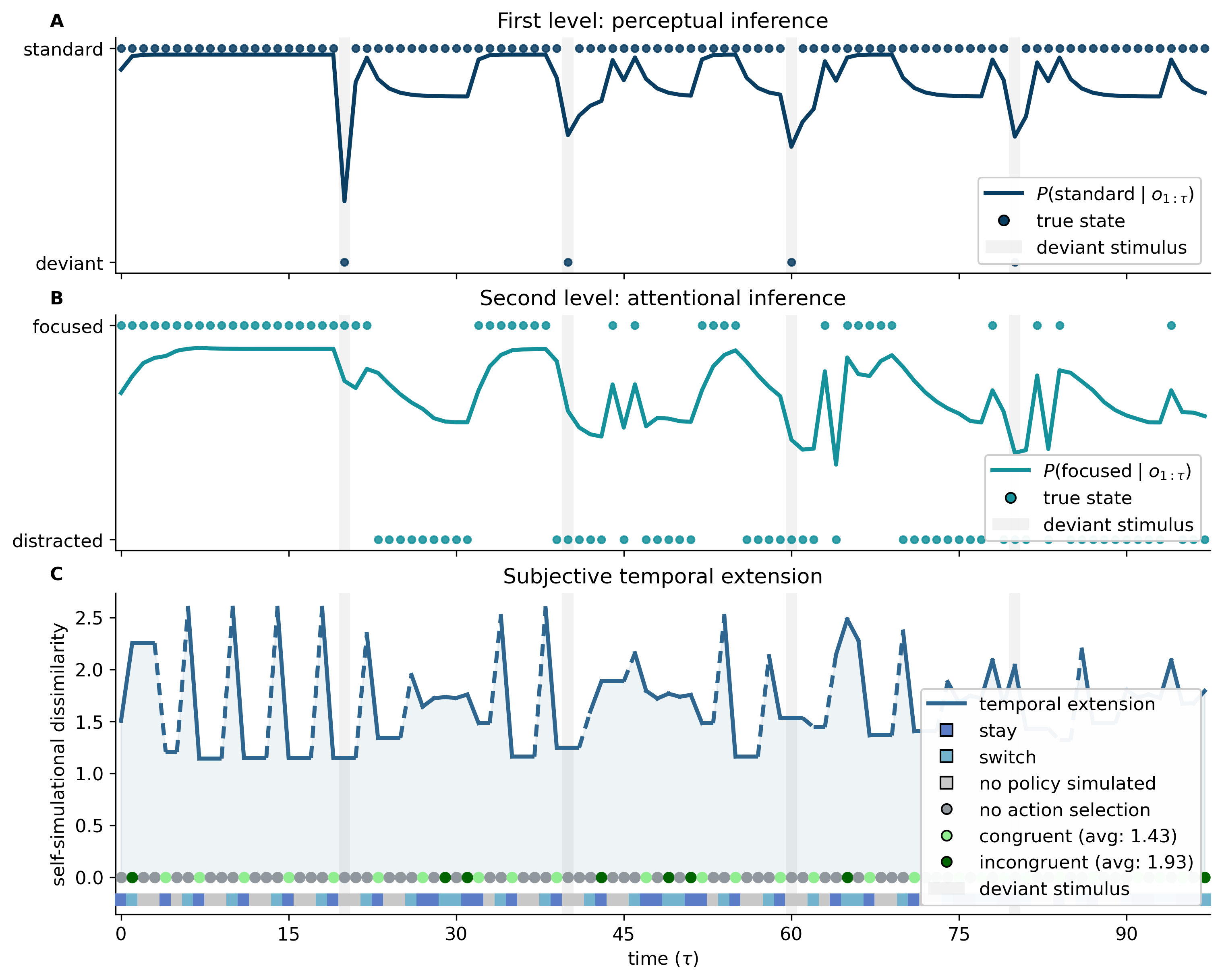}
    \caption{Simulation of a deep parametric generative model. \textbf{A} First hierarchical level: inference over 
perceptual states (‘standard’/’deviant’). \textbf{B} Second hierarchical level: inference over attentional states 
(‘focused’/’distracted’). \textbf{C} Subjective temporal extension (continuous line) and its absence (dashed line).
Coloured circles indicate simultaneity (congruency) or non-simultaneity (incongruency) between the simulation of a policy 
and its selection, if a policy is selected. Coloured boxes indicate simulated policies (‘stay’/’switch’/no policy-simulation) in the sequential simulation-epoch. Illustration inspired by (\citeauthor{sandvedsmith2021metawareness}, \citeyear{sandvedsmith2021metawareness}). Code openly available at: \url{https://github.com/JanBellingrath/deep_parametric_generative_model_of_temporal_inference}.}
    \label{fig:placeholder}
\end{figure}

\section{Variations of the ‘width’ of the subjective moment}

Psychologically, much is known concerning systematic variations of subjective time
perception. While probably everyone has experienced time slowing down in highly
emotional moments, or speeding up during pleasurable episodes of absorption,
numerous systematic variations of temporal inference have been described and
experimentally tested in detail. This empirical evidence ranges from transient situational
factors to stable dispositional constructs, including variations of temporal inference in
diverse psychopathological conditions. Methodologically, and building on current
computational (active inference) accounts of the various temporality-modulating
psychological constructs, diverse branches of empirical evidence concerning variations
of temporal inference will be explained by domain-specific modulations of the
information-theoretic structure corresponding to the ‘width’ of subjective temporality. Reports on the felt passage of time are \textit{assumed} to correspond to varying duration estimations; with a ‘faster’ felt passage of time corresponding to smaller estimated durations, and a ‘slower’ passage of time corresponding to longer estimated durations.

The extension of the subjective moment increases in negative experiences, with
temporal durations being overestimated. Experimentally, for instance, temporal
durations are overestimated when negative stimuli are expected (\citeauthor{ogden2015painanticipation} \citeyear{ogden2015painanticipation}),
fear-inducing videos are seen (\citeauthor{pollatos2014interoceptivefocus} \citeyear{pollatos2014interoceptivefocus}), or participants
recall a free-fall from a height of 31 m (\citeauthor{stetson2007freefall} \citeyear{stetson2007freefall}). Indeed, this
association is so reliable, that even the reverse inference can be empirically
demonstrated: For (illusorily) shortened time-periods, an equally intense painful stimulus
is reported to be decreased in magnitude (\citeauthor{maia2023temporalillusionpain} \citeyear{maia2023temporalillusionpain}). This emerging picture, in
which the presence of negative valence covaries with temporal overestimations, has
recently been meta-analytically supported (\citeauthor{cui2023emotiontime_meta} \citeyear{cui2023emotiontime_meta}). To give another example,
participants overestimate time when their hand is immersed in water at painful, as
opposed to neutral, temperatures, with increases in pain perception covarying with
increases in temporal overestimation (\citeauthor{rey2017paindilates} \citeyear{rey2017paindilates}). Put in the words of the authors:
“Being in pain reflects a deviation from homeostasis which threatens integrity and
prompts repeated conscious access to one’s self” (\citeauthor{rey2017paindilates} \citeyear{rey2017paindilates}). Whether it is a
painful physical sensation, a situation constraining possible behavioral options to non-desired outcomes, or whether the constraints are largely internal and self-imposed, the
dissimilarity between the outcomes expected under simulated policies (‘what could
come true, if a sequence of actions were performed’), $Q(\tilde{y}\mid \pi)$, and the outcomes
expected under prior preferences (‘what you prefer to observe’), $P(\tilde{y}\mid C)$, increases.
Indeed, negative states might be conceptually individuated by nothing but a certain
difference between expectations of possible and preferred outcomes, holding across
policy-simulations, and over a given episode, as, by definition for negative experiences, ‘you just cannot get what you want’. Essentially, if SST is correct, the ‘width’ of the subjective temporal
moment, $D_{KL}[Q(\tilde{y}\mid \pi)\Vert P(\tilde{y}\mid C)]$, is itself nothing but this difference-relation.
Accordingly, it is precisely this increased dissimilarity-relation between possible and
preferred outcomes, characterizing negative influences, that explains the increased
extension of the subjective moment in negatively-connotated circumstances; with
impressions of ‘moments feeling like eternities’ being resultant. 

This is also consistent
with a perspective offered by research on functional asymmetries in valence-processing,
documenting a deeper processing for negative stimuli across various domains
(\citeauthor{alves2017whygood} \citeyear{alves2017whygood}; \citeauthor{baumeister2001badstronger} \citeyear{baumeister2001badstronger}). Computationally, these processing
asymmetries might be manifested in terms of an increased policy-simulation frequency
and depth, necessitated because a policy evaluated in terms of sufficiently low $G(\pi)$ is
yet to be simulated – with an increased temporal ‘width’ being resultant
($D_{KL}[Q(\tilde{y}\mid \pi)\Vert P(\tilde{y}\mid C)]$ is constitutive of $G(\pi)$). Consistently, time is perceived to be
‘slower’ in depression (for a meta-analysis see \citeauthor{thoenes2015depression_meta} \citeyear{thoenes2015depression_meta}) and in
advanced cancer patients (\citeauthor{vanlaarhoven2011cancer} \citeyear{vanlaarhoven2011cancer}); and temporal overestimations
occur in schizophrenia (for a meta-analysis, see (\citeauthor{ueda2018positivesymptoms} \citeyear{ueda2018positivesymptoms})) and
PTSD (\citeauthor{vicario2018ptsdtime} \citeyear{vicario2018ptsdtime}).

While variations in valence span across many psychological domains in which temporal
processing is altered, temporal inference is also, more specifically, modulated by
impulsivity. This can be parsimoniously explained by virtue of a domain-specific variation
of the information-theoretic structure corresponding to the ‘width’ of subjective
temporality under a current active inference account of impulsivity (\citeauthor{mirza2019impulsivityactiveinference} \citeyear{mirza2019impulsivityactiveinference}).
First, empirically, a meta-analysis of temporal inference in attention-deficit
hyperactivity disorder (ADHD), a condition closely associated with increased levels of
impulsivity (e.g., \citeauthor{patros2015adhd_delaydiscounting} \citeyear{patros2015adhd_delaydiscounting}),
has shown that affected individuals overestimate
time (\citeauthor{zheng2022adhd_time} \citeyear{zheng2022adhd_time}). Borderline personality disorder, too, being also coupled to
heightened levels of impulsivity (e.g., \citeauthor{links1999impulsivitybpd} \citeyear{links1999impulsivitybpd}), is related
to a negatively felt expansion of time (\citeauthor{mioni2020bpd_time} \citeyear{mioni2020bpd_time}) and temporal overestimations
(\citeauthor{berlin2004bpd_time} \citeyear{berlin2004bpd_time}). Consistently, time is overestimated during abstinence in tobacco
use disorder (\citeauthor{miglin2017withdrawaldelay} \citeyear{miglin2017withdrawaldelay}), with time passing more slowly during withdrawal
(\citeauthor{sayette2005smokingurge} \citeyear{sayette2005smokingurge}). A similar trend has been shown in heroin users during the
withdrawal period (\citeauthor{aleksandrov2005heroin} \citeyear{aleksandrov2005heroin}). Cocaine- and methamphetamine-dependent
individuals, too, overestimate time to a degree that is correlated with their level of
impulsivity (\citeauthor{wittmann2007stimulantdependent} \citeyear{wittmann2007stimulantdependent}) (for reviews on the relationship between time
perception and impulsivity disorders, see (\citeauthor{moreira2016impulsivitydisorders} \citeyear{moreira2016impulsivitydisorders}; \citeauthor{paasche2019timeimpulsivity} \citeyear{paasche2019timeimpulsivity}); see
also (\citeauthor{wittmann2008decisionimpulsivity} \citeyear{wittmann2008decisionimpulsivity})).

Computationally, one perspective on impulsivity is in
terms of time-discounted prior-preferences (i.e., ‘selling the future for the present’) (for
an active inference account, see \citeauthor{mirza2019impulsivityactiveinference} \citeyear{mirza2019impulsivityactiveinference}). Essentially, the steeper the
discounting of the prior preferences, the stronger the preference for rewards in the
present, discarding the future (for a review on smokers devaluing the future more than
non-smokers, for instance, see (\citeauthor{barlow2017timediscounting} \citeyear{barlow2017timediscounting}); for a more general review on
temporal discounting and unhealthy behavior, see (\citeauthor{story2014tempordiscounting} \citeyear{story2014tempordiscounting})). Furthermore, the
prior preferences may not only be more strongly focused on the temporal proximity of
the rewards, but also on a specific subset of outcomes (e.g., being overly focused on
obtaining a given addictive substance, relative to other rewards). Given this ‘narrowing’ of the prior preferences, $P(\tilde{y}\mid C)$, is not mirrored (on average) by the
outcomes expected under simulated policies, $Q(\tilde{y}\mid \pi)$, the latter are (on average) more
divergent from the former. Accordingly, the ‘width’ of subjective temporality,
$D_{KL}[Q(\tilde{y}\mid \pi)\Vert P(\tilde{y}\mid C)]$, is, on average, increased for impulsive states and traits, with temporal
overestimations being resultant – thus explaining the empirical evidence.

Furthermore, the passage of time is judged to be slower in acute episodes of boredom
(e.g., \citeauthor{droitvolet2023passageoftime} \citeyear{droitvolet2023passageoftime}; see also \citeauthor{danckert2005timefliesboredom} \citeyear{danckert2005timefliesboredom}). For
example, for participants shut in an empty room alone for 7.5 minutes, the experienced
extent of boredom is associated with the feeling of time passing more slowly (\citeauthor{witowska2020waiting} \citeyear{witowska2020waiting}). Computationally, boredom has been understood in terms of switching
between exploitation and exploration: Tasks that are too easy provoke boredom,
with boredom signifying the transition from pragmatic action to novelty seeking (\citeauthor{danckert2019boredombalance} \citeyear{danckert2019boredombalance};
\citeauthor{darling2023synthesisingboredom} \citeyear{darling2023synthesisingboredom}; \citeauthor{gomezramirez2017boredomcreativity} \citeyear{gomezramirez2017boredomcreativity}; \citeauthor{parviziwayne2023forgettingourselves} \citeyear{parviziwayne2023forgettingourselves}).
Mathematically, if the pragmatic value (i.e., $\mathbb{E}_{Q(\tilde{y}\mid \pi)}[\ln P(\tilde{y}\mid C)]$) currently afforded by a
given situation, decreases, $\mathbb{TE}$ (i.e., $D_{KL}[Q(\tilde{y}\mid \pi)\Vert P(\tilde{y}\mid C)]$) increases, explaining
empirical evidence on an overly extended temporal moment in acute states of boredom.

\begin{table*}[p]
\centering
\setlength{\belowcaptionskip}{8pt} 
\caption{\textbf{Psychological effect classes computationally explained by SST}
}
\label{tab:sst_effect_classes}
\setlength{\tabcolsep}{6pt}
\renewcommand{\arraystretch}{1.25}
\normalsize

\setlength{\extrarowheight}{3.1pt} 

\begin{tabularx}{\textwidth}{@{}p{0.22\textwidth} p{0.20\textwidth} Y@{}}
\toprule
\textbf{\textit{Effect class}} & \textbf{\textit{Empirical finding}} & \textbf{\textit{Mechanism}} \\
\midrule

\textbf{Agency (intentional binding)}
& Contraction / underestimation
&  The comparatively low expected free energy of policies corresponding to subsequently enacted actions maps onto a low \(\mathbb{TE}
\), thus producing the intentional binding effect. \\
\\
\textbf{Meditation / minimal phenomenal experience}
& 'Atemporality'/ 'timeless change'
& Reduced frequency of counterfactual policy simulation implies that \(\mathbb{TE}
\) is transiently not (or less) constructed; perceptual inference can still evolve over time, yielding dynamic succession without temporal extension ('timeless change'). \\
\\
\textbf{Attention-to-time effects}
& Dilation / overestimation
& Precision as attention:  Attention to time, \textit{is}, by SST, simply attention to the (weak) first person perspective; which, if outcomes under policies, \(Q(\tilde y\mid \pi)\), do not similarly contract, implies an increased temporal extension. \\
\\
\textbf{Central tendency effects}
& Regression-to-mean bias
& A direct consequence of Bayesian duration inference (i.e., the influence of the duration inference prior). \\
\\

\textbf{States and dispositions: Negative valence / threat / pain}
& Dilation / overestimation
& Increased mismatch between outcomes expected under simulated policies \(Q(\tilde y\mid \pi)\) and phenotypically preferred outcomes \(P(\tilde y\mid C)\) increases \(D_{\mathrm{KL}}\!\big[Q(\tilde y\mid \pi)\Vert P(\tilde y\mid C)\big]\). \\

\\
\textbf{Dispositional impulsivity (ADHD, borderline)}
& Dilation / overestimation
& Time-discounted and/or narrowed prior preferences increase \(\mathbb{TE}
\) \\
\\
\textbf{Boredom}
& Dilation / overestimation
& A comparatively reduced pragmatic value in a given situation mathematically implies an increased \(\mathbb{TE}
\). \\
\\
\textbf{Flow}
& Contraction / underestimation
& A comparatively high pragmatic value mathematically implies a comparatively small \(\mathbb{TE}
\). \\
\\
\textbf{Addiction / withdrawal}
& Dilation / overestimation
& Narrowed, time-discounted preferences and frustrated goal-states increase the divergence \(D_{\mathrm{KL}}[Q(\tilde y\mid \pi)\Vert P(\tilde y\mid C)]\). 
\\

\\
\bottomrule
\end{tabularx}
\end{table*}

On the other hand, the flow-state (\citeauthor{csikszentmihalyi1990flow} \citeyear{csikszentmihalyi1990flow}), characterized both by an increase
in concentration on a task that is subjectively perceived as meaningful and challenging,
and by a decrease of narrative self-modelling, has repeatedly been associated with a
faster temporal experience (e.g., \citeauthor{im2018distortedflowtime} \citeyear{im2018distortedflowtime}; \citeauthor{rutrecht2021thumperflow} \citeyear{rutrecht2021thumperflow}). Indeed, participants even reverse infer having been in
the flow-state, given that they are (falsely) informed about much time having passed
(\citeauthor{christandl2018timeflows} \citeyear{christandl2018timeflows}). Computationally, a recent active inference
account explains the attenuation of narrative self-modelling in flow-experiences by
virtue of an exclusive attentional focus mandated by the challenging nature of the task
(\citeauthor{parviziwayne2023forgettingourselves} \citeyear{parviziwayne2023forgettingourselves}). Furthermore, by this account, because the agent is, in the
flow-state, because of experience and training, already confident about the
consequences of actions, action-selection is primarily influenced by pragmatic (as
opposed to epistemic) value. Accordingly, and again because a high pragmatic value
mathematically maps onto a small $\mathbb{TE}$ (i.e., a low $D_{KL}[Q(\tilde{y}\mid \pi)\Vert P(\tilde{y}\mid C)]$), this explains
the faster temporal experience empirically attested in the flow-state.

Consider another well-known phenomenon: Paying attention to time leads to longer estimations of durations, a classic result described as
“the most robust, well-replicated finding in the time perception literature” (for a review
see \citeauthor{brown2010_time_review} \citeyear{brown2010_time_review}). This effect is often studied in terms of dual-task interference tasks,
wherein the additional presence of a non-timing task, concurrently performed with a
timing task, leads to decreases of the estimated durations, interpreted in terms of a
relatively decreased attention allocated to the timing task. Consistently, if the difficulty
of the non-timing task increases, the magnitude of the interference effect increases (e.g.,
\citeauthor{brown1985_dualtask} \citeyear{brown1985_dualtask}; \citeauthor{zakay1983_dualtask} \citeyear{zakay1983_dualtask}), and the interference effect reduces if
the participants are trained in either one of the two tasks (e.g., \citeauthor{brown_bennett2002_training} \citeyear{brown_bennett2002_training}). Indeed, a review on the family of dual-task interference effects (i.e., smaller
estimated durations, higher variability \& higher inaccuracy) found that 91\% out of 77
studies conducted between 1924 and 2008 showed this interference effect (\citeauthor{brown2008_dualtaskreview} \citeyear{brown2008_dualtaskreview}; see also \citeauthor{brown1997_timeattention} \citeyear{brown1997_timeattention}).

Paying increased attention to time
entails, under the usual mapping of precision to attention, and the perspective offered by SST, a transiently increased precision of prior preferences: If temporal extension is a consequence of self-modelling, anchored on the first person perspective (see section 3), attention to time simply \textit{is} attention to the first person perspective (for a generalized active inference model of mental actions, see \citeauthor{sandvedsmith2021metawareness} \citeyear{sandvedsmith2021metawareness}).
If, however, the precision of prior preferences increases without a corresponding contraction of the outcomes expected under counterfactual policies, $D_{KL}[Q(\tilde{y}\mid \pi)\Vert P(\tilde{y}\mid C)]$ increases. 
Thus, SST explains
why increased duration estimates ensue when attention is being paid to time (for a
review on attentional effects on time see (\citeauthor{brown2008_dualtaskreview} \citeyear{brown2008_dualtaskreview}); for the classic attentional-gate model, often referenced as an
explanation in this context, see (\citeauthor{zakay_block1997_gate} \citeyear{zakay_block1997_gate})).

As a last empirical constraint, consider near-death experiences. In near-death 
experiences, such as car-crashes or other accidents, the subjective temporal Now is
reported to be extended to such a degree that time subjectively ‘stands still’ (\citeauthor{noyes1976depersonalization} \citeyear{noyes1976depersonalization}). Indeed, how could the predicted outcomes of a simulated policy be more
divergent from prior preferences (with the subjective temporal Now, $\mathbb{TE}$, extending
accordingly) than the expected outcomes of a policy where one is predicted to be dead,
or potentially so?

\section{Conclusion}

Nothing in the physics of time corresponds to an extended moment in the subjective
sense. This paper has described how the ‘width’ of subjective temporality emerges as a transparent
and counterfactual inference from the dynamics of self-modelling, cast in terms of free
energy minimization. Intuitively, because a given counterfactual self-simulation could, if selected, be instantiated in the three- (and \textit{n}-) dimensional space of bodily (and mental) actions,
and because it is associated with a dissimilarity-relation, of how far it diverges from this
instantiation, it extends in a distinct dimension of self-simulational counterfactuality,
anchored on these spaces. But this transparent counterfactual extension simply is what
an inference of subjective temporal ‘width’ conceptually presupposes – hence the
proposed mechanism. 
Here, this self-simulational dissimilarity relation was expressed as an \textit{actively inferred} temporal extension, and cast 
in the parlance of information geometry. Following a simulation of a deep parametric generative model illustrating the emergence of two empirically attested alterations of temporal inference (i.e., the
intentional binding effect \& alterations of temporal experience in deep meditative
states), numerous systematic variations of temporal experience and biases of temporal
inference have been explained via the proposed mathematical structure. This concerns
classic psychological effects such as attentional and central-tendency effects, as well as variations of subjective temporal experience
across varying levels of valence, boredom, impulsivity, flow-states, near-death experiences, and various psychopathologies, amongst others. Across all these constructs, the
explanatory strategy has been the same: Demonstrating that current active inference
accounts mapping onto the respective constructs mathematically imply systematic
variations of \textit{precisely} that information-theoretic structure that is -- by SST – identical to
subjective temporal ‘width’ in \textit{precisely} the direction that the psychological evidence
respectively indicates. This systematic covariation between the dynamics of self-modelling and temporal inference is consistent with neuroimaging meta-analyses
revealing both self-modelling and temporal inference as centrally implicated in insula
functioning (for a meta-analysis on self-modelling, see e.g., \citeauthor{qin2020threelevelself}, \citeyear{qin2020threelevelself}; for meta-analyses on temporal inference, see (\citeauthor{mondok2023selectivitytiming}, \citeyear{mondok2023selectivitytiming}; \citeauthor{naghibi2023embodyingtime}, \citeyear{naghibi2023embodyingtime})). The proposed self-simulational dissimilarity-relation is epistemically
parsimonious: not only because the computation of a dissimilarity-relation between
‘what could’ and ‘what ought’ is psychologically all-prevalent but also because the
corresponding mathematical structure is implied by the minimization of expected free energy. Concerning states of minimal phenomenal
experience (i.e., awareness of nothing but awareness) (\citeauthor{metzinger2024elephantblind}, \citeyear{metzinger2024elephantblind}), entirely
\textit{atemporal} (without a subjective temporal Now) and \textit{non-dual }(without a phenomenal
self-model), SST describes how atemporality emerges as a consequence of non-duality, thus potentially paving a way towards a deeper understanding of minimal phenomenal
experience itself.

“\textit{I measure myself as I measure time}” (\citeauthor{heidegger1992historyconcepttime}, \citeyear{heidegger1992historyconcepttime})

\section*{Acknowledgements}
I am grateful to Daniel Friedman, Karl Friston, Thomas Metzinger, Lars Sandved-Smith, Marc Wittmann and members of the Minimal Phenomenal Experience Project for their constructive comments and feedback. The code is open-source and can be accessed under: 
\url{https://github.com/JanBellingrath/deep_parametric_generative_model_of_temporal_inference}


\bibliographystyle{plainnat}
\bibliography{references}

\end{document}